\documentclass[10pt,conference,compsocconf,letterpaper]{IEEEtran}
\pdfoutput=1

\usepackage{times, fullpage, amsmath, amsfonts, amssymb, url, xspace, graphics}
\IEEEoverridecommandlockouts 

\newcommand{\remove}[1]{}
\newcommand{\removeinfinal}[1]{} % final 

\remove{
\newtheoremstyle{coolstyle}
    {9pt}% Space above
    {9pt}% Space below
    {\slshape}% Body font
    {}% Indent amount
    {\bfseries}% Theorem head font
    {.}% Punctuation after theorem head
    {.5em} % Space after theorem head
    {}% Theorem head spec (can be left empty, meaning `normal')
\theoremstyle{coolstyle}
}

\newtheorem{theorem}{Theorem}[section]

\newtheorem{lemma}[theorem]{Lemma}

\newcommand{\Id}[1]{\textit{#1}}

\newcommand{\IF}{\textbf{if}\xspace}
\newcommand{\ELSE}{\textbf{else}\xspace}
\newcommand{\FE}{\textbf{foreach}\xspace}

\newcommand{\WHILE}{\textbf{while}\xspace}
\newcommand{\RET}{\textbf{return}\xspace}
\newcommand{\BREAK}{\textbf{break}\xspace}
\newcommand{\OUT}{\textbf{output}\xspace}
\newcommand{\BECOMES}{$:=$\xspace}
\newcommand{\T}{\hspace{1em}}

\newcommand{\TT}{\T\T}
\newcommand{\TTT}{\T\T\T}
\newcommand{\TTTT}{\T\T\T\T}
\newcommand{\TTTTT}{\T\T\T\T\T}

\newcommand{\ws}{\mathit{ws}}

\newcommand{\RC}{\mathit{RC}}

\newcommand{\Op}{\mathit{op}}

\newcommand{\OpA}{\Op_1}
\newcommand{\OpB}{\Op_2}

\newcommand{\Stime}{s}
\newcommand{\Ftime}{f}

\newcommand{\Proof}{\noindent \emph{Proof:}\ \ }

\newcommand{\FZ}{\mathit{FZ}}
\newcommand{\BZ}{\mathit{BZ}}

\newcommand{\bp}{\mathit{bp}}
\newcommand{\spt}{\mathit{sp}}

\newcommand{\zstart}{\underline{f}}
\newcommand{\zfinish}{\bar{s}}

\begin{document}

\title{On the $k$-Atomicity-Verification Problem}
\author{
  \thanks{$^{\textstyle \dag}$ Authors completed part of this research at HP Labs.}
  \thanks{$^{\textstyle \ddag}$ Author partially supported by the Natural Sciences and Engineering Research Council (NSERC) of Canada.}
  \IEEEauthorblockN{Wojciech Golab $^{\textstyle \dag \ddag}$}
  \IEEEauthorblockA{Department of Electrical and Computer Engineering\\University of Waterloo, wgolab@uwaterloo.ca}
  \and
  \IEEEauthorblockN{Jeremy Hurwitz \quad Xiaozhou (Steve) Li $^{\textstyle \dag}$}
  \IEEEauthorblockA{Google\\\{hurwitz, xzli\}@google.com}
}

\maketitle

% !TEX root = main.tex
% !TEX spellcheck = en_US

\begin{abstract}

Modern Internet-scale storage systems often provide weak consistency
in exchange for better performance and resilience.  An important weak
consistency property is $k$-atomicity, which bounds the staleness of
values returned by read operations.  The $k$-atomicity-verification
problem (or $k$-AV for short) is the problem of deciding whether a
given history of operations is $k$-atomic.  The 1-AV problem is
equivalent to verifying atomicity/linearizability, a well-known and
solved problem.  However, for $k \geq 2$, no polynomial-time $k$-AV
algorithm is known.

This paper makes the following contributions towards solving the
$k$-AV problem.  First, we present a simple 2-AV algorithm called LBT,
which is likely to be efficient (quasilinear) for histories that arise
in practice, although it is less efficient (quadratic) in the worst
case.  Second, we present a more involved 2-AV algorithm called FZF,
which runs efficiently (quasilinear) even in the worst case.  To our
knowledge, these are the first algorithms that solve the 2-AV problem
fully.  Third, we show that the weighted $k$-AV problem, a natural
extension of the $k$-AV problem, is NP-complete.

\end{abstract}

% \newpage
	
% !TEX root = main.tex
% !TEX spellcheck = en_US

\section{Introduction}
\label{sec:intro}

Data consistency is an important consideration in storage systems.
Modern Internet-scale storage systems often provide weak (rather than
strong) consistency in exchange for better performance and resilience.
An important weak consistency property is
\emph{$k$-atomicity}~\cite{aab:k-consist}.  A history of operations is
called \emph{$k$-atomic} iff there exists a valid total order on the
operations (i.e., one that conforms to the partial order imposed by
the operation time intervals) such that every read obtains one of the $k$
freshest values with respect to that total order.  By this definition,
the well-known
atomicity/linearizability~\cite{hw:linear,lamport:registers,misra:axioms}
is equivalent to 1-atomicity.
%When $k>1$, $k$-atomicity allows a read to obtain a stale value; the smaller the $k$ the stronger the consistency.

The $k$-atomicity property is well-suited to describing the behavior
of replicated storage systems that employ non-strict (i.e., ``sloppy'')
quorums \cite{aab:k-consist}, such as Amazon's Dynamo.
In such systems, reads may return stale values because read and write quorums
are not guaranteed to overlap.
% cite Dynamo
Classic consistency properties such as safety and regularity~\cite{lamport:registers} fail
to capture this behavior, and instead require that reads return the freshest value,
except in the special case when they overlap with a write.
%Another advantage of $k$-atomicity is its simplicity, which makes it
%suitable for specifying the behavior or storage systems.
%Furthermore, Lamport's safety property allows reads to return arbitrary values
%if they overlap with a write, whereas $k$-atomicity ensures that reads only
%return values that were previously written.
Furthermore, many modern applications can tolerate $k$-atomicity very well.
For example, in a social network, a user may still be satisfied that,
although the data retrieved is not the latest, it is at most a few
updates behind.

Verifying that a storage system satisfies a certain consistency
property serves two purposes.  First, we would like to know whether
a system delivers what it promises in terms of consistency, as
theoretically correct storage protocols can have buggy
implementations.  Second, we would like to know whether a system
provides more consistency than is needed for a particular
application, making it possible to turn back certain ``tuning knobs''
(e.g., quorum size) and reduce operational costs.

The $k$-atomicity-verification problem (or $k$-AV for short) is to
decide whether a given history is $k$-atomic.  The 1-AV problem (also
called verifying linearizability) is well-known and
solved~\cite{kvs-team:hotdep,gk:tsm,misra:axioms}.  However,
for $k>1$, no polynomial-time $k$-AV algorithm is known, except that Golab,
Li, and Shah~\cite{golab+ls:podc11} solved the case $k=2$ for a
restricted class of histories.

This paper makes the following contributions towards solving
the $k$-AV problem.  First, we present a simple 2-AV algorithm called
LBT for arbitrary histories (Section~\ref{sec:lbt}).  LBT's simplicity
makes it attractive for implementation.  Although it is quadratic in
the worst case, it is likely to be quasilinear for the common cases
that arise in practice.  Second, we present a more involved 2-AV algorithm
called FZF for arbitrary histories
(Section~\ref{sec:fzf}).  FZF is more efficient in the worst case as
it always runs in quasilinear time.  To our knowledge, LBT and FZF are
the first algorithms that fully solve the 2-AV problem.  Third, we
prove that the weighted $k$-AV problem, a natural generalization of
the $k$-AV problem, is NP-complete (Section~\ref{sec:k-wav}).

% LocalWords:  read's write's linearizability iff Aiyer et al VL Golab WAV

% !TEX root = main.tex
% !TEX spellcheck = en_US

\section{Model}
\label{sec:model}

\subsection{Terminology and notations}
\label{sec:model-terms}

We consider a storage system that supports read and write operations,
where each operation runs for some finite amount of time.  Operations on
different storage locations are independent of each other, and so we can
model the storage system as a collection of read/write
registers~\cite{lamport:registers,misra:axioms}.  A \emph{history} is
a collection of operations on the same register, where each operation
has a start time, finish time, type (read or write), and value
(retrieved or stored).  Let the start time and finish time of an
operation $\Op$ be $\Op.\Stime$ and $\Op.\Ftime$ respectively.  We say
that $\OpA$ \emph{precedes} $\OpB$ (and $\OpB$ \emph{succeeds}
$\OpA$), denoted $\OpA < \OpB$, iff $\OpA.\Ftime < \OpB.\Stime$.
If neither $\OpA < \OpB$ nor $\OpB < \OpA$, then $\OpA$ and $\OpB$ are
\emph{concurrent} with each other.  For a read, its \emph{dictating
  write} is the unique write whose value the read obtains.  For
a write, the reads that obtain its value are called the
\emph{dictated reads} of the write.  A write can have zero or more
dictated reads.

The ``precedes'' relation defines a partial order.  A total order of
the operations is called \emph{valid} if it conforms to this partial
order. 
% consider removing ``commit point'' later
Equivalently, a total order is valid iff there exists a distinct point
within the time interval of each operation, called the \emph{commit point} (where the
operation appears to take effect), such that the order of the commit
points determines the total order~\cite{hw:linear}.
A valid total order is called \emph{$k$-atomic} iff, in this total
order, every read follows its dictating write and is separated from
this write by at most $k-1$ other writes.  A history is called
\emph{$k$-atomic} iff it has a valid $k$-atomic total order.

\subsection{Problem statement}

Given a history, we would like to decide whether the history is
$k$-atomic, where $k$ is a given value of interest (typically a small
constant).  We call this problem the $k$-atomicity-verification
problem (or $k$-AV for short).
% WG: the following sentence repeats something from the intro
%By this definition, the 1-AV problem
%is equivalent to the problem of verifying atomicity/linearizability,
%which has been solved in the
%literature~\cite{kvs-team:hotdep,gk:tsm,misra:axioms}.  
Given a solution to $k$-AV for arbitrary $k$, we can use
binary search to compute the smallest $k$ for which a history is $k$-atomic.
Note that like atomicity, $k$-atomicity is a local property~\cite{hw:linear},
and so we can solve $k$-AV by reasoning independently about
each register accessed in a history.

\subsection{Assumptions}
\label{sec:model-assume}

We assume that each write assigns a distinct value, for two reasons.
First, in our particular practical application (storage systems), all
writes can be tagged with a globally unique identifier, for example
consisting of the local time of a machine issuing the write followed
by the machine's identifier. Therefore, this assumption does not incur
any loss of generality.  Second, if the values written are not unique,
then the decision problem of verifying consistency properties is
NP-complete for several well-known properties, in particular
1-atomicity and sequential
consistency~\cite{cls:complex,gk:tsm,taylor:complex}.  We further
assume that all start times and finish times are unique.

We assume also that the values written/retrieved are integers and that the
start and finish of each operation can be timestamped accurately.
Recent work in systems has made it feasible for timestamps to closely reflect 
real time, even in a highly distributed environment.
For example, the TrueTime API in Spanner provides highly
accurate estimates of real time~\cite{corbett:spanner}.  The
operations under consideration in this paper typically last for tens
or hundreds of milliseconds, whereas a clock can be read in 
approximately 100 microseconds.  Therefore, we ignore the potential
overhead of reading clock values.

By the definition of $k$-atomicity, a history may contain anomalies
that immediately prevent it from being $k$-atomic.  These anomalies
are: a read without a dictating write, or a read that precedes its
dictating write.  Detection of such anomalies is straightforward and
we assume that the given history does not contain them.

Lastly, we assume that a write ends before any of its dictated reads.
If a given history does not satisfy this assumption, we can enforce it
by shortening writes so that their finish time is slightly smaller than
the minimum finish time of their dictated reads.
We do so without loss of generality because a write's commit point
cannot occur after one of its dictated reads has finished.

% LocalWords:  iff Korach

% !TEX root = main.tex
% !TEX spellcheck = en_US

\section{The LBT algorithm}
\label{sec:lbt}

In this section, we present the first 2-AV algorithm LBT, which uses a
technique called \emph{limited backtracking}~\cite{even+is:timetable};
hence the name.  It is exceedingly simple and is likely to run in
nearly linear time in practice.

%---------------------------------------------------------------------

\subsection{The algorithm}
\label{sec:lbt-alg}

\begin{figure}[tbp]
  \center{\scalebox{0.7}{\includegraphics{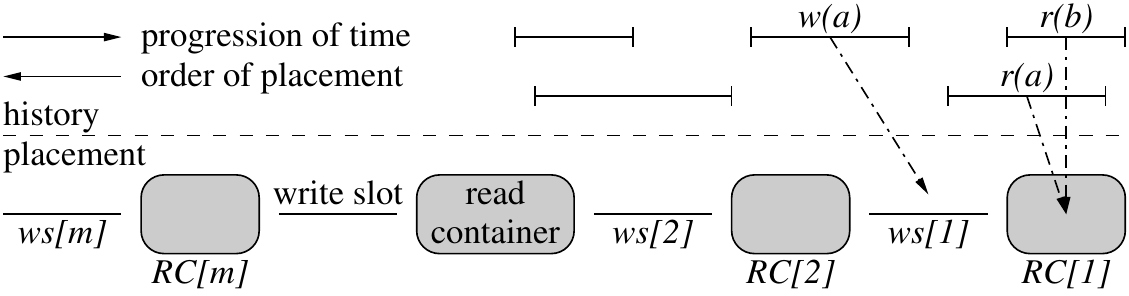}}}
  \caption{Placing operations into write slots and read containers.}
  \label{fig:slots-containers}
\end{figure}

Conceptually, LBT attempts to construct a 2-atomic total order.  It
examines operations in the given history from back to front and places
them into a sequence of write slots and read containers, where each
write slot holds exactly one write and each read container holds zero
or more reads.  The resulting total order is defined by the order of
the write slots and read containers.  The order of the reads in the
same read container is unimportant (and as such, left unspecified) as
long as they conform to the ``precedes'' partial order.  See
Figure~\ref{fig:slots-containers} for an illustration of write slots
and read containers.

LBT runs in epochs.  At the beginning of each epoch, it tentatively
puts a candidate in the latest unfilled write slot, say $\ws[i]$.
This first placement determines the reads to be placed into the
adjacent read container $\RC[i]$, and what goes into $\RC[i]$ then
determines what goes into $\ws[i-1]$. This then determines $\RC[i-1]$,
and so on.  An epoch ends when a read container placement does not
constrain the subsequent write slot placement.  If during the run of
an epoch, some placement cannot be satisfied, LBT aborts the current
epoch and considers a different candidate as the first write in this
epoch.  LBT outputs NO if all candidates are exhausted.
Figure~\ref{alg:lbt} presents the LBT algorithm.  Although the code in
Figure~\ref{alg:lbt} does not explicitly maintain the write slots and
read containers, it is not hard to see that lines 18 and 20 correspond
to placing the operations into them.

It is important to note that, once the first write of an epoch is
placed, the rest of the writes in the epoch as well as their order are
uniquely determined: no further search is necessary in this epoch.
Furthermore, backtracking is limited to the start of an epoch.  These
two properties ensure the efficiency of the algorithm.

%---------------------------------------------------------------------

%---------------------------------------------------------------------

\begin{figure}[tbp]
\begin{tabbing}
1 $H$ \BECOMES original history; $W$ \BECOMES all the writes in $H$;\\
2 \WHILE ($H \neq \emptyset$) \{\\
3 \T $C$ \BECOMES all the writes in $W$ that do not\\
  \TTTTT precede any other writes in $W$;\\
4 \T \FE ($w \in C$) \{\\
5 \TT \Id{success} \BECOMES RunEpoch($w$, $H$, $W$);\\
6 \TT \IF (\Id{success}) \BREAK;\\
7 \TT revert $H$ and $W$ to before RunEpoch; \}\\
8 \T \IF (not \Id{success}) \OUT NO; \}\\
9 \OUT YES;\\
\\
\hspace{1.75em}// all parameters are in/out parameters\\
10\ \ bool RunEpoch($w$, $H$, $W$) \{\\
11 \T \WHILE (true) \{\\
12 \TT $w'$ \BECOMES $\bot$;\\
13 \TT \FE ($\Op \in H$ where $w.f < \Op.s$) \{\\
14 \TTT \IF ($\Op$ is write) \RET false;\\
15 \TTT \IF ($\Op$'s dictating write $\neq w$ and $\neq w'$) \{\\
16 \TTTT \IF ($w' \neq \bot$) \RET false;\\
17 \TTTT $w'$ \BECOMES $\Op$'s dictating write; \}\\
18 \TTT $H$ \BECOMES $H \setminus \{\Op\}$; \}\\
19 \TT $R$ \BECOMES remaining dictated reads of $w$ in $H$;\\
20 \TT $H$ \BECOMES $H \setminus R \setminus \{w\}$; $W$ \BECOMES $W \setminus \{w\}$;\\
21 \TT \IF ($w'$ is $\bot$) \RET true;\\
22 \TT $w$ \BECOMES $w'$; \}\ \}
\end{tabbing}
  \caption{The LBT algorithm.}
  \label{alg:lbt}
\end{figure}

%---------------------------------------------------------------------

\subsection{Correctness}
\label{sec:lbt-correct}

% !TEX root = main.tex
% !TEX spellcheck = en_US

% \section{Proof of correctness for the LBT algorithm (Section~\ref{sec:2-av-lbt})}
% \label{sec:proof-lbt}

The main intuition behind the correctness of LBT is that (1) LBT tries
all possible candidates at each epoch, and (2) once an epoch succeeds,
the remaining history is 2-atomic iff the original history is.

\begin{theorem}
  \label{thm:lbt-correct}
  LBT outputs YES iff $H$ is 2-atomic.
\end{theorem}

\IEEEproof We first show that if LBT outputs YES, then the original
history is 2-atomic.  Given a YES execution of LBT, consider the total
order induced by the operations placed into write slots (line 20) and
read containers (lines 18 and 20).  To see that this total order
always conforms to the ``precedes'' partial order, consider two
operations $\Op_1$ and $\Op_2$ where $\Op_1 < \Op_2$.  Suppose $\Op_1$
is a write.  If $\Op_2$ is a write, then $\Op_2$ cannot be one of the
operations in line 13 when $w = \Op_1$, otherwise line 14 would have
rejected the epoch.  If $\Op_2$ is a read, then because $\Op_1 <
\Op_2$, line 18 ensures that $\Op_2$ is placed before $\Op_1$.  Now
suppose $\Op_1$ is a read.  One way that $\Op_1$ can be placed into a
read container is on line 18, in which case $\Op_2$ would have to be
placed into the same or earlier read container due to the condition on
line 13.  Another way is on line 20.  Since we assume that $w.f <
\Op_1.f$ (i.e., a write's finish time is less than any of its dictated
read's; see Section~\ref{sec:model-assume}), we have $w.f < \Op_2.s$
because $\Op_1 < \Op_2$, implying that $\Op_2$ is placed no later than
$\Op_1$ due to the condition on line 13.

In the total order constructed, for each read, either $w$ is the
dictating write, resulting in no intervening writes, or $w'$ is set to
the dictating write (line 17), resulting in one intervening write.
Therefore, the total order is 2-atomic, and so is the original
history.

It remains to show that if the original history is 2-atomic, then LBT
outputs YES.  Note that, if the original history is 2-atomic and if
RunEpoch succeeds, then the remaining history is 2-atomic. This is
because adding more operations to a non-2-atomic history only keeps it
so.  Since LBT tries all possible candidates in each epoch, it will
find a proper candidate that makes RunEpoch succeed, and we can repeat
the above argument for the remaining history, which is again 2-atomic.
\IEEEQED

% LocalWords: LBT

%---------------------------------------------------------------------

\subsection{Time complexity}
\label{label:lbt-time}

Let $n$ be the total number of operations in the original history and
let $c$ be the maximum number of concurrent writes at any time.  We do
not assume that the operations in the original history are sorted.

\begin{theorem}
  \label{lbt-time}
  LBT can be implemented to run in $O(n \log n + c \cdot n)$ time.
\end{theorem}

\IEEEproof We maintain $H$ as a doubly linked list sorted by start
time, and $W$ as a doubly linked list sorted by finish time.  For each
write $w$, we maintain a doubly linked list of all of $w$'s dictated
reads; for each of these reads, we add a pointer to point to its
counterpart in $H$.  In $H$, we add two pointers from a read to its
dictating write in $H$ and $W$.  In $W$, we add a pointer from a write
to its list of dictated reads.  Clearly, all the pre-processing above
takes $O(n \log n)$ time.

Since $W$ is sorted by finish time, identifying the writes in $C$ on
line 3 takes $O(c)$ time as they form a suffix of $W$.  In the search
for a successful candidate for an epoch, we have to try $O(c)$
candidates.  If implemented as described in Figure~\ref{alg:lbt}, we
may run into the situation where a successful candidate is examined
late, while early candidates take a long time to fail.

We can use the technique of iterative deepening~\cite{iter-deep} with
the search depth doubled in each iteration to make the examination of
candidates run faster. In each iteration $i$, for all the candidates
in $C$, we execute RunEpoch to length $2^i$. As soon as one candidate
returns true, we declare this epoch successful and discard all other
candidates.  Executed this way, the execution of lines 4--7 takes
total running time $O(c \cdot t)$, where $t$ is the time taken to find
the shortest successful candidate if there is one, or $t$ is the time
needed for the last surviving candidate to return false if there is
no successful candidate.

Because of the data structures outlined at the beginning of this
proof, identifying all the operations on line 13 takes $O(c)$ time,
removing an $\Op$ from $H$ on line 18 takes constant time, identifying
$R$ on line 19 takes constant time, removing $R$ from $H$ on line 20
takes $|R|$ time, and removing $w$ from $W$ takes constant time.
Therefore, $t$ is proportional to the number of operations removed
from $H$.  Since $|H|=n$, the running time of LBT (after
pre-processing) is $O(c \cdot n)$.  \IEEEQED

Theoretically, $c$ can be as high as $n$, so LBT's worst-case running
time is $O(n^2)$.  However, in practice, most applications only have a
small number of concurrent writes at any time.  Therefore, we believe
that, helped by its simplicity, LBT will run very well in practice.

% LocalWords:  LBT pre iff

% !TEX root = main.tex
% !TEX spellcheck = en_US

\section{The FZF algorithm}
\label{sec:fzf}

In this section we present another 2-AV algorithm called Forward Zones First (FZF).
Before we describe the algorithm and explain its name,
%The name of the algorithm is derived from the order in which it considers
%the operations in the input history, which are organized 
%of forward zones, and then deals with backward zones separately.
%Compared to LBT, FZF is more involved, but also more efficient in the worst case.
%Internally, FZF uses a simplified variant of LBT as a subroutine.
%---------------------------------------------------------------------
%
%\subsection{Additional terminology}
%\label{sec:fzf-terms}
we first review some terminology introduced by Gibbons and
Korach~\cite{gk:tsm}.  A \emph{cluster} is a subset of operations
in a history that comprises a write and its
dictated reads.  The \emph{zone} $Z$ for a cluster is the time interval
between the minimum finish time of any operation in the cluster, denoted
by $Z.\zstart$, and the maximum start time of any such operation in the cluster, denoted
by $Z.\zfinish$.  A zone $Z$ is called a \emph{forward zone} if
$Z.\zstart < Z.\zfinish$, otherwise it is called a \emph{backward zone}.
The \emph{low} endpoint of $Z$, denoted by $Z.l$, is
$\min(Z.\zstart, Z.\zfinish)$.  The \emph{high} endpoint of $Z$,
denoted by $Z.h$, is $\max(Z.\zstart, Z.\zfinish)$. 
Using these definitions, Gibbons and Korach~\cite{gk:tsm} show that a history is 1-atomic if and only if:
(1) no two forward zones overlap, and
(2) no backward zone is contained entirely in a forward zone.

As its name suggests, algorithm FZF processes the input history
by considering forward zones before backward zones.
Note that in describing FZF, we will discuss clusters and zones interchangeably,
for example by referring to forward and backward clusters.

\subsection{The algorithm}
\label{sec:fzf-algo}
It is natural to ask whether the 2-AV problem can be solved using an
approach that, like Gibbons and Korach's 1-AV algorithm, considers
only the set of forward and backward zones corresponding to a history.
The answer to this question is negative as it is possible to construct two
histories, one 2-atomic and the other not, that have identical sets of zones
\cite{golab+ls:podc11}.
Consequently, a 2-AV algorithm must analyze the history at a deeper
level than by looking at zones alone.

FZF is a three-stage algorithm that breaks up an input history into smaller chunks in
Stage~1, and then analyzes each chunk separately in Stage~2.
In Stage~1, chunks are chosen on the basis of zones only.
Stage~2 then considers additional details of the history while analyzing
each chunk.
For each chunk, the algorithm attempts to construct a 2-atomic total order
over its operations by first ordering the dictating writes corresponding to
forward zones, and then dealing with backward zones.
Finally, in Stage~3 the input history is deemed 2-atomic if and only if each chunk 
considered in Stage~2 is 2-atomic.
As we explain later on in Section~\ref{fzf-correct}, FZF is
correct under the assumptions stated in Section~\ref{sec:model}.

In this section we first describe the algorithm informally, and then present
pseudo-code in Figure~\ref{alg:fzf}.
%We now describe each stage in more detail, letting $H$ denote the input history.

\newcommand{\chunks}{\ensuremath{\mathcal{CS}}}

\medskip
\noindent\textit{Stage~1}\\
In order to define Stage~1 precisely, we first introduce some additional terminology and notation.
A \emph{chunk} of the input history $H$ is a set of clusters in $H$ such that:
\begin{enumerate}
\item the union of forward zones for these clusters is a continuous
and non-empty time interval, and
\item the union of backward zones corresponding to these clusters is a subset
of the former interval.
\end{enumerate}
The \emph{projection of $H$ onto a chunk $K$}, denoted $H|K$, is 
the subhistory $H$ that contains all the operations for clusters in $K$.
A chunk is called \emph{maximal} if adding another
cluster to it breaks one of the above two properties.
Next, we define the \emph{chunk set of $H$}, denoted $\chunks(H)$,
as the set of maximal chunks of $H$ such that every forward cluster in $H$
belongs to some chunk in the set.
Finally, we call a cluster \emph{dangling} if it does
not belong to any chunk in $\chunks(H)$.
It follows directly from the definition of $\chunks(H)$ that every dangling cluster
is a backward cluster.

\begin{figure}[htbp]
  \center{\scalebox{0.75}{\includegraphics{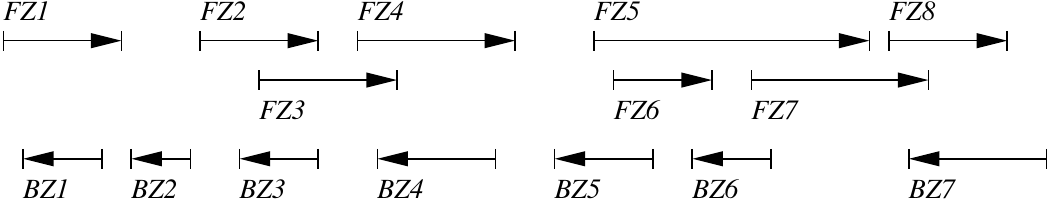}}}
  \caption{Example illustrating Stage~1 of FZF.
The algorithm identifies three maximal chunks:
$\left\{\FZ_1, \BZ_1\right\}$, $\left\{\FZ_2, \FZ_3, \FZ_4, \BZ_3, \BZ_4\right\}$, and
$\left\{\FZ_5, \FZ_6, \FZ_7, \FZ_8, \BZ_6\right\}$.
There are also three dangling clusters, corresponding to $\BZ_2$, $\BZ_5$ and $\BZ_7$.}
\remove{ (a) three chains of
    forward zones after stage 1; we have deliberately included the
    back-stretch point and low endpoint for $\FZ_6$; (b)
    ``back-stretching'' of forward zone $\FZ_6$ in stage 2; we have
    drawn the back-stretch point, the schedule point, and the low
    endpoint; (c) considering backward zones in stages 3 and 4; we
    have drawn the $t_{1,2,3}$ points of time for all three chains.}
  \label{fig:fzf}
\end{figure}

Stage~1 of FZF simply computes $\chunks(H)$ from $H$, thus breaking the input into
smaller pieces similarly to a divide-and-conquer algorithm.
(Note, however, that FZF does not divide the pieces recursively.)
For example, given a history that yields the zone structure shown in Figure~\ref{fig:fzf},
Stage~1 identifies three maximal chunks and three dangling clusters, as explained
in the caption.

% Careful: there may be two viable orders on the forward zone writes that need to be considered

\medskip
\noindent\textit{Stage~2}\\
The goal of Stage~2 is to decide for each maximal chunk $K \in \chunks(H)$ whether or not
$H|K$ is 2-atomic by testing carefully chosen orderings over dictating writes of $K$.
Given a subset $S$ of such dictating writes, and a candidate total order $T_S$ over them,
FZF uses a subroutine to verify that $T_S$ is \emph{viable},
meaning that there exists a valid 2-atomic total order over the writes in $S$ and their dictated reads.
If $S$ contains all of the dictating writes in $K$, then the existence of a viable total order $T_S$
over $S$ implies that $H|K$ is 2-atomic.
As we explain later on in Section~\ref{sec:fzf-space-time},
a subroutine that checks whether $T_S$ is viable can be implemented
using a simplified version of the LBT algorithm from Section~\ref{sec:lbt}.

To find a viable total order on all the dictating writes of $K$, FZF
first considers the subset of writes in forward clusters
(hence the algorithm's name).
Two such total orders are considered:
$T_F$, which orders writes in increasing order of the low endpoints of their forward zones; and
$T_F'$, which is constructed from $T_F$ by swapping the first two writes.
(If $T_F$ has only one element then $T_F = T_F'$.)
We prove in Section~\ref{fzf-correct} (see Lemma~\ref{lemma:fzf-stage2-a})
that it suffices to consider only these two orders.

Next, the algorithm incorporates the dictating writes of backward clusters.
As we show in Section~\ref{fzf-correct} (see proof of Lemma~\ref{lemma:fzf-stage2-b}),
if $K$ contains three or more backward clusters then $H|K$ is not
2-atomic, hence FZF outputs NO and terminates.
Otherwise, FZF considers extensions of $T_F$ and $T_F'$ where the dictating writes of backward clusters
are either appended or pre-pended, at most one write at either end.
For example, if the dictating writes of backward zones are $w_1$ and $w_2$, then the
algorithm considers $w_1 T_F w_2$, $w_2 T_F w_1$, $w_1 T_F' w_2$, and $w_2 T_F' w_1$.
As we show in Section~\ref{fzf-correct} (see Lemma~\ref{lemma:fzf-stage2-b}),
it suffices to consider only these four orders to decide whether $H|K$ is 2-atomic.
Similarly, up to four total orders are considered if $K$ contains only one backward cluster, and
up to two (i.e., $T_F$ and $T_F'$ themselves) if $K$ has no backward clusters.
For each total order chosen, FZF invokes the subroutine described earlier
%(i.e., simplified LBT algorithm) 
to decide if the total order is viable.
If none of the total orders is viable, FZF outputs NO and terminates.

\medskip
\noindent\textit{Stage~3}\\
If the algorithm reaches Stage~3, then each chunk considered in Stage~2 was deemed 2-atomic.
As we show in Section~\ref{fzf-correct} (see Lemma~\ref{lemma:subdivide}), this
implies that $H$ is 2-atomic, hence the algorithm outputs YES and terminates.

\begin{figure}[tbp]
\begin{tabbing}
// input: history $H$ \\
// output: YES if $H$ is 2-atomic, NO otherwise \\
%$\epsilon :=$ min distance between any two operation endpoints;\\
%construct the zones for the given history $H$;\\
// Stage 1\\
compute chunk set $\chunks(H)$;\\
%compute dangling clusters $D_1, D_2, D_3, \ldots$ in $H$; \\
// Stage 2\\
%construct the disjoint chains for the forward zones;\\
\FE (chunk $K \in \chunks(H)$) \{\\
\T $T_F := $ sequence of writes in $K$, in increasing order \\
\T\T\T of the low endpoints of their forward zones; \\
\T $T_F' := $ $T_F$ with first two elements interchanged\\
\T\T\T ($T_F' := T_F$ if $T_F$ only has one element);\\
%\T \IF ((three fwd.\ zones in $K$ overlap at one point) or\\
%\T (one fwd.\ zone in $K$ overlaps three others)) \{\\
\T $B := $ number of backward clusters in $K$;\\
\T \IF ($B = 0$) \{\\
\T\T $\mathcal{S} := \{T_F,  T_F'\};$\\
\T \} \ELSE \IF ($B = 1$) \{\\
\T \T $w := $ dictating write of the backward cluster;\\
\T\T $\mathcal{S} := \{w T_F,  T_F w, w T_F', T_F' w\};$\\
\T \} \ELSE \IF ($B = 2$) \{\\
\T \T $w_1,w_2 = $ dictating writes of the backward clusters;\\
\T\T $\mathcal{S} := \{w_1 T_F w_2,  w_2 T_F w_1, w_1 T_F' w_2, w_2 T_F' w_1\};$\\
\T \} \ELSE \IF ($B \geq 3$) \{\\
\T \T  $S := \emptyset$; \T // $H$ is definitely not 2-atomic \\
\T \} \\
\T \FE (total order $T \in S$) \{ \\
\T \T  use subroutine to check if $T$ is viable; \}\\
\T \IF ($S = \emptyset$ or none of the total orders in $S$ is viable) \{ \\
\T \T \OUT NO and terminate; \} \}\\
// Stage 3\\
\OUT YES; \T // $H$ is 2-atomic \\
\end{tabbing}
  \vspace{-12pt}
  \caption{The FZF algorithm.}
  \label{alg:fzf}
\end{figure}

%---------------------------------------------------------------------

\subsection{Correctness}
\label{fzf-correct}

% !TEX root = main.tex
% !TEX spellcheck = en_US

% \section{Proof of correctness for the FZF algorithm}
% \label{sec:proof-fzf}

It follows easily that FZF terminates, and so it suffices to show that
the algorithm outputs YES if and only if the given history $H$ is 2-atomic.
%If FZF outputs YES, it also assigns the
%schedule points to the zones, which implies the assignments of the
%commit points.  Then our obligation is to show that the total order
%induced by these commit points is indeed 2-atomic.  If FZF outputs NO,
%then our obligation is to show that in each case where FZF outputs NO,
%it is indeed impossible for $H$ to be 2-atomic.
We reach this goal through a sequence of technical lemmas.
The first lemma captures the rationale behind dividing the input into
maximal chunks in Stage~1:

\begin{lemma}
  \label{lemma:subdivide}
  For every history $H$, $H$ is 2-atomic if and only if for every chunk
  $K \in \chunks(H)$, $H|K$ is 2-atomic.
\begin{proof} Suppose that $H$ is 2-atomic, and let $T$ be a valid 2-atomic total order
on the operations in $H$.
For any $K \in \chunks(H)$, let $T|K$ be the total order over operations
in $H|K$ such that $T$ extends $T|K$.
Then $T|K$ is a valid 2-atomic total order on the operations in $H|K$.
%and hence $H|K$ is 2-atomic.

Conversely, suppose that for every chunk $K \in \chunks(H)$, $H|K$ is 2-atomic.
We will show how to construct a valid 2-atomic total order on the operations in $H$.
%\newline\underline{Case 1:} $H$ has no dangling clusters.
%Then the set of operations in $H$ is the union of the sets of operations corresponding
%to the clusters in $\chunks(H)$.
For each $K \in \chunks(H)$, let $K.l$ denote the minimum $Z.l$ for any zone $Z$
corresponding to a cluster in $K$, and let $K.h$ denote the maximum $Z.h$ 
for any such $Z$.
%Given distinct chunks $K,K' \in \chunks(H)$, let $K <_H K'$ denote that $K.h < K'.l$,
%and note that $<_H$ is a total order since each chunk in $\chunks(H)$ is maximal.
Next, for each $K \in \chunks(H)$, let $T_K$ denote a valid 2-atomic total order on the
operations in $H|K$.
Similarly, for each dangling cluster $D_j$ in $H$, define a valid 2-atomic total order $T_{D_j}$ over operations in $D_j$.
(The existence of $T_{D_j}$ follows from assumptions stated in Section~\ref{sec:model}.)
Also let $D.l$ and $D.h$ denote $Z.l$ and $Z.h$, respectively, where $Z$ is the zone corresponding
to cluster $D$.

Now define the relation $\leq_H$ over chunks and dangling clusters as follows:
given elements $X,Y$, each either a chunk in $\chunks(H)$ or a dangling cluster of $H$,
$X \leq_H Y$ denotes that $X.h < Y.l$.
It follows easily that $\leq_H$ is a partial order.
Furthermore, since all chunks in $\chunks(H)$ are maximal, it follows that all
such chunks are totally ordered by $\leq_H$.
Finally, choose any total order that extends $\leq_H$, and let $T$ denote the concatenation
in that order of all $T_{K_i}$ and $T_{D_j}$ for each chunk $K_i \in \chunks(H)$ and 
each dangling cluster $D_j$ of $H$.

It follows easily that $T$ is a 2-atomic total order on all the operations in $H$.
It remains to show that $T$ is valid (i.e., extends the ``precedes'' relation over operations in $H$).
Suppose for contradiction that $T$ is not valid.
Then there exist distinct operations $Op,Op'$ as well as two elements $X,Y$,
each either a chunk in $\chunks(H)$ or a dangling cluster of $H$,
such that $Op$ belongs in $X$, $Op'$ belongs in $Y$,
$T_X$ precedes $T_Y$ in $T$, and yet $Op'$ precedes $Op$ in $H$.
(In this context, ``belongs'' means that an operation is either part of a cluster in some maximal chunk,
or is part of some dangling cluster.)
\newline\underline{Case~1:} $X$ and $Y$ are both chunks.
Since $T_X$ precedes $T_Y$ in $T$ and since $T$ totally orders all chunks in $\chunks(H)$,
it follows that $X \leq_H Y$ holds, and hence $X.h < Y.l$.
Next, note that $Op$ starts no later than $X.h$, otherwise the zone for some cluster in $X$ would extend to after $X.h$,
and similarly $Op'$ finishes no earlier than $Y.l$, otherwise the zone for some cluster in $Y$ would extend to before $Y.l$.
Since $X.h < Y.l$, this implies that $Op$ starts before $Op'$ finishes.
But that contradicts $Op'$ preceding $Op$.
\newline\underline{Case~2:} $X$ and $Y$ are both dangling clusters.
Since $T_X$ precedes $T_Y$ in $T$, $Y \leq_H X$ is false, and so $Y.h \geq X.l$.
%Since we assume in Section~\ref{sec:model} that the start and finish times of all operations in $H$ are unique,
%this implies that $Y.h > X.l$.
Next, note that since $X$ and $Y$ are both backward clusters,
$X.l$ is a point inside $Op$ and $Y.h$ is a point inside $Op'$.
Since $Y.h \geq X.l$, this implies that $Op'$ finishes no earlier than $Op$ starts.
But that contradicts $Op'$ preceding $Op$ in $H$.
\newline\underline{Case~3:} $X$ is a chunk and $Y$ is a dangling cluster, and $T_X$ precedes $T_Y$ in $T$.
Since $T_X$ precedes $T_Y$ in $T$, $Y \leq_H X$ is false, and so $Y.h \geq X.l$.
Furthermore, since $Y$ is a dangling cluster, $Y$ is not part of chunk $X$, and so $Y.h > X.h$.
Next, note that $Op$ starts no later than $X.h$, otherwise the zone for some cluster in $X$ would extend to after $X.h$.
Also, since $Y$ is a backward cluster, $Y.h$ is a point inside $Op'$.
Since $Y.h > X.h$, this implies that $Op'$ finishes after $Op$ starts.
But that contradicts $Op'$ preceding $Op$ in $H$.
\newline\underline{Case~4:} $X$ is a dangling cluster and $Y$ is a chunk, and $T_X$ precedes $T_Y$ in $T$.
The proof is analogous to Case~3.
We show that $Y.l > X.l$, hence $Op'$ finishes after $Op$ starts, which contradicts $Op'$ preceding $Op$ in $H$.
\end{proof}
\end{lemma}

\medskip
Next, we show the correctness of Stage~2 of FZF.
To simplify presentation, we introduce a definition:
letting $T$ denote a valid total order on a subset $S$ of the writes in $H$, we say that 
the \emph{separation of a write $w \in S$ in $T$} is the minimum number of writes
that separate $w$ from any of its dictated reads (not including $w$ itself) in any valid total order $T'$ that
extends $T$ to both the writes in $S$ and their dictated reads.
It follows that if $T$ is viable, then the separation of every $w \in S$ in $T$
is less than two.

\begin{lemma}
\label{lemma:fzf-stage2-a}
For any chunk $K \in \chunks(H)$, if an iteration of the outer for loop occurs in Stage~2 for $K$, then
any viable total order $T$ over the writes of forward clusters in $K$ is an element
of $\{T_F, T_F'\}$, where $T_F$ and $T_F'$ are computed at the beginning of the iteration.
\begin{proof}
Suppose that $T$ does exist, and note that the pattern of forward zones in $K$
cannot have the following property, denoted $P$ in the remainder of the proof:
three zones overlap at one point, or one zone overlaps more than two others.
(If $K$ has property $P$ then one can show that in any valid total order $T'$ over the writes in $K$,
 the dictating write for one of the forward zones has separation at least two in $T'$, and hence $T'$ is not viable.)
As a result, each forward zone overlaps at most two others, forming a ``chain''
of forward zones resembling one of the three shown in Figure~\ref{fig:fzf}.
Note that since $K$ is a maximal chunk, this chain has no ``breaks'' in it,
and so all but two of the zones overlap exactly two others.
% "all but two" means "all but two or fewer"
We proceed by induction on the number of forward clusters in $K$, denoted $f$ in this proof.
If $f \leq 2$ then the set $\{T_F, T_F'\}$ contains all the possible total orders under consideration,
and so the lemma holds.
Now suppose that the lemma holds for some $f \geq 2$, and consider a chunk with $f+1$ forward clusters.
Label the forward zones of these clusters as $A, B, C, \ldots$ in increasing order of their low endpoints,
and let $w_A, w_B, w_C, \ldots$ denote the corresponding dictating writes.
Note that $T_F = w_A, w_B, w_C, \ldots$ and $T_F' = w_B, w_A, w_C, \ldots$.

\noindent\underline{Case~1:} $A$ ends before $B$ ends.
Since $K$ does not have property $P$, the chain of zones in this case initially resembles the
middle chunk in Figure~\ref{fig:fzf}, with $A = \FZ_2$, $B = \FZ_3$, and $C = \FZ_4$.
Next, note that if cluster $A$ were removed from chunk $K$, then by the induction hypothesis
any viable total order on $w_B,w_C,\ldots$ would be either $T_F'' = w_B,w_C,\ldots$
or $T_F''' = w_C,w_B,\ldots$.
% Note: technically we may have to remove some backward clusters in addition to $A$, otherwise
% $K$ becomes something that's not a chunk when we remove $A$.
For chunk $K$, this implies that $T$ must order the writes for $B, C, \ldots$ in the same manner as either $T_F''$ or $T_F'''$.
Now consider the possible positions of $w_A$ in $T$.
For each case, we must either show that $T$ is identical to $T_F$ or $T_F'$, or else derive a contradiction
by showing that $T$ is not viable.
%$T_F = w_A, w_B, w_C, w_D, \ldots$ or $T_F' = w_B, w_A, w_C, w_D, \ldots$.
\newline\noindent\underline{Subcase~1a:} $w_A$ is the second or later element in $T$.
Then $T = w_B, \ldots, w_A, \ldots$ (if $T$ extends $T_F''$) or $T = w_C, \ldots, w_A, \ldots$ (if $T$ extends $T_F'''$).
In the first case, since $w_A$ and $w_C$ both precede some dictated read of $w_B$ in $H$,
$w_B$ has separation at least two in $T$, and so $T$ is not viable.
In the second case, since $w_A$ and $w_B$ both precede some dictated read of $w_C$ in $H$,
$w_C$ has separation at least two in $T$, and so $T$ is not viable.
\newline\noindent\underline{Subcase~1b:} $w_A$ is the first element in $T$.
Then $T = w_A, w_B, w_C, \ldots$ (if $T$ extends $T_F''$) or $T = w_A, w_C, w_B, \ldots$ (if $T$ extends $T_F'''$).
In the first case, $T$ is identical to $T_F$.
In the second case, since $w_B$ precedes some dictated read of $w_A$ in $H$,
$w_A$ has separation at least two in $T$, and so $T$ is not viable.

\noindent\underline{Case~2:} $A$ ends after $B$ ends.
Since $K$ does not have property $P$, the chain of zones in this case initially resembles the
rightmost chunk in Figure~\ref{fig:fzf}, with $A = \FZ_5$, $B = \FZ_6$, and $C = \FZ_7$.
We proceed as in Case~1, but invoke the induction hypothesis on 
$A,C, \ldots$ instead of $B,C,\ldots$.
We deduce that $T$ must order the writes for $A,C, \ldots$ in the same manner as either
$T_F'' = w_A,w_C,\ldots$ or $T_F''' = w_C,w_A,\ldots$.
Next, we consider the possible positions of $w_B$ in $T$.
\newline\noindent\underline{Subcase~2a:} $w_B$ is the second or later element in $T$.
Then $T = w_A, \ldots, w_B, \ldots$ (if $T$ extends $T_F''$) or $T = w_C, \ldots, w_B, \ldots$ (if $T$ extends $T_F'''$).
In the first case, since $w_B$ and $w_C$ both precede some dictated read of $w_A$ in $H$,
$w_A$ has separation at least two in $T$, and so $T$ is not viable.
In the second case, since $w_A$ and $w_B$ both precede some dictated read of $w_C$ in $H$,
$w_C$ has separation at least two in $T$, and so $T$ is not viable.
%is separated from one of its dictated reads by both $w_B$ and $w_C$.
\newline\noindent\underline{Subcase~2b:} $w_B$ is the first element in $T$.
Then $T = w_B, w_A, w_C, \ldots$ (if $T$ extends $T_F''$) or $T = w_B, w_C, w_A, \ldots$ (if $T$ extends $T_F'''$).
In the first case, $T$ is identical to $T_F'$.
In the second case, since $w_A$ precedes some dictated read of $w_B$ in $H$,
$w_B$ has separation at least two in $T$, and so $T$ is not viable.
\end{proof}
\end{lemma}

\begin{lemma}
\label{lemma:fzf-stage2-b}
For any chunk $K \in \chunks(H)$, if an iteration of the outer for loop occurs in Stage~2 for $K$, then
any viable total order $T$ over all the writes of $K$ is an element
of the set $S$ computed in this iteration.
\begin{proof}
Suppose that $T$ exists.
It follows from Lemma~\ref{lemma:fzf-stage2-a} that $T$ must order the writes in $K$
consistently with either $T_F$ or $T_F'$, which are computed at the beginning of the iteration,
otherwise $T$ is not viable.
Next, note that in both $T_F$ and $T_F'$, each write except the last one has separation one,
otherwise either $T$ is not viable (if separation is higher than one) or $K$ is not a
maximal chunk (if separation is zero).
Since $T$ extends either $T_F$ and $T_F'$, this implies that
the dictating writes of any backward zones in $K$ cannot be placed in $T$ between
two dictating writes of forward zones, otherwise the separation of one of the latter writes
becomes greater than one, and hence $T$ is no longer viable.
In other words, the dictating writes of backward zones must be placed in $T$ either before
or after all the writes of forward zones.
We use this observation in the case analysis below.

Let $B$ denote number of backward clusters in $K$.

\noindent\underline{Case~1: $B = 0$}.
Then $T$ must be either $T_F$ or $T_F'$, and indeed the algorithm includes both
orders in $S$.

\noindent\underline{Case~2: $B = 1$}.
Let $w$ denote the dictating write of the backward cluster, as in the algorithm.
Then $T$ must be either append or pre-pend $w$ to either $T_F$ or $T_F'$.
Indeed the algorithm includes all four possible orders in $S$.

\noindent\underline{Case~3: $B = 2$}.
Let $w_1,w_2$ denote the dictating writes of the backward clusters, as in the algorithm,
and let $B_1,B_2$ denote the corresponding backward zones.
Then in $T$, $w_1$ must either precede or follow all the writes of forward clusters,
and similarly for $w_2$.

We show first that $w_1$ and $w_2$ in $T$ cannot both precede all the writes of forward clusters.
Let $w_f$ be the dictating write of the forward cluster in $K$ whose forward zone has the earliest low endpoint,
and let $F$ denote this zone.
Suppose for contradiction that $T$ places $w_1$ and $w_2$ before $w_f$, in that order (without loss of generality).
Since $K$ is a maximal chunk, $B_1.l > F.l$ holds, which means that some operation in $B_1$
starts after some operation in $F$ ends.
As a result, any valid 2-atomic total order $T'$ over $H|K$ that extends $T$ places some operation of $B_1$ after $w_f$.
($T'$ exists because we assume that $T$ is viable and exists.)
Since we assume that $T$ places $w_1$ before $w_f$, $T'$ must place some dictated read $r_1$ of $w_1$ after $w_f$.
In that case $w_1$ is separated from $r_1$ by both $w_2$ and $w_f$.
Thus, $w_1$ has separation at least two in $T$,
which contradicts $T$ being viable.

Next, we show that $w_1$ and $w_2$ in $T$ cannot both succeed all the writes of forward clusters.
Let $w_f$ be the dictating write of the forward cluster in $K$ whose forward zone has the largest high endpoint,
and let $F$ denote this forward zone.
Suppose for contradiction that $T$ places $w_1$ and $w_2$ after $w_f$, in that order (without loss of generality).
Since $k$ is a maximal cluster, $B_1.h < F.h$ holds, which means that some operation in $B_1$
ends before some dictated read $r_f$ in $F$ begins.
(Recall that every forward cluster has at least one dictated read.)
As a result, any valid 2-atomic total order $T'$ over $H|K$ that extends $T$ places some operation of $B_1$ before $r_f$.  
(Again, $T'$ exists because we assume that $T$ is viable and exists.)
Since $T'$ is valid and 2-atomic, this implies that $T'$ places $w_1$ before $r_f$.
By an analogous argument, $T'$ places $w_2$ before $r_f$.
Since $w_1$ and $w_2$ both follow $w_f$ in $T$, this implies that $w_f$ has separation at least two in $T$,
which contradicts $T$ being viable.

Thus, $T$ can only be one of four possible total orders:
$w_1 T_F w_2$, $w_2 T_F w_1$, $w_1 T_F' w_2$, and $w_2 T_F' w_1$.
Indeed the algorithm includes all four of these in $S$.

\noindent\underline{Case~4: $B \geq 3$}.
Let $w_1,w_2,w_3, \ldots$ denote the dictating writes of the backward clusters.
Then in $T$, each of these writes must either precede or follow all the writes of forward clusters.
This contradicts our observation from Case~3 that at most one of $w_1,w_2,w_3, \ldots$ can precede,
and at most one can follow, all the writes of forward clusters.
\end{proof}
\end{lemma}

\begin{lemma}
\label{lemma:fzf-stage2-c}
For any chunk $K \in \chunks(H)$, if an iteration of the outer for loop occurs in Stage~2 for $K$,
and this iteration outputs NO, then $H|K$ is not 2-atomic.
Conversely, if the iteration for chunk $K$ occurs and does not output NO, then $H|K$ is 2-atomic.
\begin{proof}
Suppose that an iteration of the outer for loop occurs in Stage~2 for $K$, computes the set $S$, and outputs NO.
Since the algorithm outputs NO, none of the total orders in $S$ are viable, and hence by 
Lemma~\ref{lemma:fzf-stage2-b} there is no viable total order $T$ over the writes of all clusters in $K$.
This implies that $H|K$ is not 2-atomic.

Conversely, suppose that the iteration for chunk $K$ occurs and does not output NO.
Then some total order $T \in S$ is viable.
Furthermore, it follows from the algorithm for Stage~2 that $T$ is a total order
over the writes of all clusters in $K$.
Since $T$ is viable, this implies that $H|K$ is 2-atomic.
\end{proof}
\end{lemma}

Finally, the following theorem asserts the overall correctness of FZF:

\begin{theorem}
\label{theorem:fzf-stage2}
For any input history $H$, if $H$ is 2-atomic then algorithm FZF outputs YES, otherwise it outputs NO.
\begin{proof}
Suppose first that FZF outputs YES.
Then the outer for loop iterates over all the chunks in Stage~2 without outputting NO, and so
by Lemma~\ref{lemma:fzf-stage2-c}, $H|K$ is 2-atomic for each maximal chunk $K \in \chunks(H)$.
Then by Lemma~\ref{lemma:subdivide}, $H$ itself is 2-atomic, as wanted.
Otherwise, suppose that FZF outputs NO.
Then this occurs in Stage~2, and so by Lemma~\ref{lemma:fzf-stage2-c} there is some chunk
$K \in \chunks(H)$ such that $H|K$ is not 2-atomic.
This implies that $H$ is not 2-atomic either, as wanted.
\end{proof}
\end{theorem}

\subsection{Time complexity}
\label{sec:fzf-space-time}
Let $n$ denote the number of operations in $H$.

\begin{theorem}
Algorithm FZF can be implemented to run in $O(n \log n)$ time. % and $O(n)$ space.
\begin{proof}
Suppose that $H$ is given as a sequence of events in arbitrary order.
The algorithm can perform the following pre-processing in $O(n \log n)$ steps before Stage~1:
create a mapping $M$ from values to clusters using a balanced tree data structure, and represent
each cluster as a linked list of operations;
then for each cluster identify its zone;
and for each zone record the value assigned by its dictating write,
its low and high endpoint, and its type (i.e., forward or backward).

In Stage~1 the algorithm can compute $\chunks(H)$ by iterating over the clusters
in $M$, inserting zones into an interval tree sorted by the low zone endpoint,
and finally iterating over the zones to identify maximal chunks.
Chunk $K$ can be represented as a list $L_K$ of dictating writes of clusters in $K$.
To simplify Stage~2, the writes in $K$ can be sorted in the same order as their
zones in the interval tree, and also tagged with the corresponding zone type.
Finally, $\chunks(H)$ can be represented as a linked list of pointers to chunks.
Thus, Stage~1 can be performed in $O(n \log n)$ steps:
$O(n \log n)$ to build $M$,
$O(n \log n)$ to traverse the balanced tree underlying $M$ and build the interval tree representing zones,
$O(n)$ to traverse the interval tree and compute maximal chunks, and finally $O(n)$ to record $\chunks(H)$.

In Stage~2, the outer for loop iterates over each maximal chunk $K$ by walking a linked list
representation of $\chunks(H)$.
This list traversal takes $O(n)$ steps, and furthermore the algorithm performs work for each
chunk $K \in \chunks(H)$.
Now let $n_K$ denote the number of operations in chunk $K$.
Aside from the inner for loop, the body of the outer for loop for chunk $K$ computes $T_F$,
which is obtained easily in $O(n_K)$ steps from the representation of $K$ described earlier.
$T_F'$ can then be obtained in $O(n_K)$ steps from $T_F$.
Similarly, the dictating writes of backward zones can be identified from the representation of $K$
and counted in $O(n_K)$ steps.
The algorithm then computes up to four total orders by pre-pending or appending
up to two dictating writes of backward zones to $T_F$ and $T_F'$, which takes $O(n_K)$ steps.

Finally, consider the inner for loop, which iterates over up to four total orders over the writes of
all clusters in chunk $K$, and tests whether each order is viable.
To test whether such an order $T$ is viable, it suffices to first check if $T$ is valid,
which takes $O(n_K)$ steps since $T$ has length $O(n_K)$, and
then call a subroutine to test whether $T$ can be extended to a valid 
2-atomic total order $T'$ over all the operations in $H|K$.
The latter test can be carried out using a simplified LBT algorithm (see Section~\ref{sec:lbt}) that
accepts $T$ and $H|K$ as part of its input, where a list of operations in $H|K$ can be obtained
from $M$ in $O(n_K \log n)$ time.
The simplified LBT algorithm attempts to find $T'$ by processing writes in reverse order of $T$, without back-tracking,
and for each write deciding which read operations must follow it.
Using a simplified version of the analysis from Section~\ref{label:lbt-time}, it follows
that this takes $O(n_K \log n_K)$ steps.
Thus, in Stage~2 of algorithm FZF, the iteration of the outer for loop for chunk $K$ runs in 
$O(n_K \log n)$ steps.

Since Stage~3 merely outputs YES, it follows that algorithm FZF can be implemented as described above
to run in $O(n\log n)$ steps in total.
%Furthermore, the data structures described in the above analysis require $O(n)$ space.
\end{proof}
\end{theorem}

\remove{
\subsection{Optimizations}
\label{sec:fzf-opti}

{\Huge TODO}
}

% LocalWords:  FZF commit linearizes tuples tuple linearizing versa FZ BZ LBT
% LocalWords:  linearize

% !TEX root = main.tex
% !TEX spellcheck = en_US

\section{The weighted $k$-AV problem ($k$-WAV)}
\label{sec:k-wav}

In this section, we show that a natural extension of the $k$-AV
problem, called the \emph{weighted $k$-AV} problem (or $k$-WAV for
short), is NP-complete.  The $k$-WAV problem is defined similarly to
the $k$-AV problem, except that each write comes with an positive
integer weight, and the $k$-atomicity requirement is that the
total weight of the writes separating any dictating write from any of
its dictated reads (including the dictating write itself) is at most
$k$.  By this definition, the $k$-AV problem is a special case of the
$k$-WAV problem where each write has weight equal to 1.  The $k$-WAV
problem captures the notion of ``important'' writes, which can have a
higher weight than ``unimportant'' writes: a read can be intervened
from its dictating write by a larger number of unimportant writes, but
only a smaller number of important writes.  A storage system can
potentially mark certain write operations to be important and require
that they not be separated from their dictated reads by too many other
important writes.

% !TEX root = main.tex
% !TEX spellcheck = en_US

% \section{Proof of NP-hardness for $k$-WAV (Section~\ref{sec:k-wav})}
% \label{sec:proof-k-wav}

\begin{theorem}
  \label{thm:k-wav}
  The $k$-WAV problem is NP-complete.
\end{theorem}

\Proof It is straightforward to see that $k$-WAV is in NP.  To prove
that $k$-WAV is NP-hard, we reduce from the well-known bin-packing
problem~\cite{gj:np-book}.  In the bin-packing problem, we are given a
set of $n$ items, each with a size $s_i$ that is a positive integer
for $1 \leq i \leq n$, a bin capacity $B$, and $m$ bins.  We are
asked whether there is a partition of these $n$ items into $m$
disjoint sets such that the sum of the sizes of the items in each
subset is at most $B$, the bin capacity.

\begin{figure}[tbp]
  \center{\scalebox{0.65}{\includegraphics{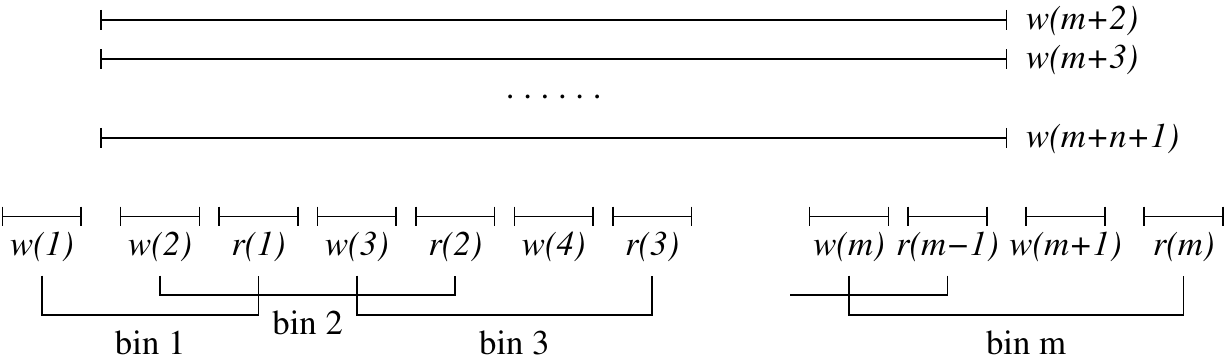}}}
  \caption{Construction for the NP-completeness of $k$-WAV.}
  \label{fig:k-wav}
\end{figure}

Given an instance of the bin-packing problem, we construct an instance
of the $k$-WAV problem as shown in Figure~\ref{fig:k-wav}.  In the
figure, it is understood that all end points are slightly different to
follow our assumption that all end points have a distinct timestamp.
In the figure, the $n$ ``long writes'' have weights equal to the sizes of the $n$ given
items in the bin-packing problem instance.  The $m$ ``short writes''
each have weight 1.  The intervals are constructed in such a way that
the short writes and their dictated reads are totally ordered, that
is, $w(1) w(2) r(1) w(3) r(2) w(4) r(3) \dots
w(m) r(m-1) w(m+1) r(m)$.  Thus, solving the $k$-WAV instance is tantamount
to deciding the commit points of the long writes, which have to occur after $w(1)$ and before
$w(m+1)$.  The total weight of the long writes placed between $w(i)$
and $r(i)$ is bounded by $B$.  In other words, we are setting $k=B+2$
for the $k$-WAV problem.  We note that the long writes do not have
dictated reads and so they can be placed anywhere between $w(1)$ and
$w(m+1)$ provided that they observe the bin capacity limit between $w(i)$ and
$r(i)$ for $1 \leq i \leq m$.  The short write $w(m+1)$ is a ``dummy''
write so as to ensure that bin $m$ (from $w(m)$ to $r(m)$)
has available capacity $B$ (but not $B+1$) for the long writes.

We now prove that the bin-packing problem instance has a solution iff
the $k$-WAV problem has a solution.  If the bin-packing problem
instance has a solution, then for those items that go into bin 1 we
place the corresponding long writes between $w(1)$ and $w(2)$.  For
those items that go into bin $i$, where $2 \leq i \leq n$, we place
the corresponding long writes between $r(i-1)$ and $w(i+1)$.  This
placement satisfies both the validity requirement and the
$(B+2)$-atomicity requirement of the $k$-WAV problem.

On the other hand, if the $k$-WAV problem instance has a solution,
then we can construct a solution for the bin-packing problem as
follows.  We first observe that in the solution for the $k$-WAV
problem, if there is a long write that is placed between $w(i)$ and
$r(i-1)$ where $2 \leq i \leq m$, then we can always re-place this
long write to between $r(i-1)$ and $w(i+1)$ and the new solution is
still a valid solution for the $k$-WAV problem instance.  This is
because placing a long write in the former manner increases the
``load'' on two bins: $i-1$ (from $w(i-1)$ to $r(i-1)$) and $i$
(from $w(i)$ to $r(i)$).
Yet placing a long write in the latter manner only
increases the load on bin $i$.  Therefore, we can
always transform a solution for the $k$-WAV problem instance so that
no long write places load on multiple bins.  Then we can
straightforwardly convert the solution to the $k$-WAV problem instance
to a solution for the bin-packing problem instance.  \QED

% LocalWords: WAV iff

%---------------------------------------------------------------------

% LocalWords:  WAV

% !TEX root = main.tex
% !TEX spellcheck = en_US

\section{Related work}
\label{sec:related}
Our 2-AV algorithms are inspired by several prior results.
The $k$-AV problem for $k=1$ has been solved in prior work on specifying
and verifying shared memories~\cite{gk:tsm,misra:axioms}.
A partial solution to the 2-AV problem appears in~\cite{golab+ls:podc11}.
The technique of limited backtracking used in our LBT algorithm (Section~\ref{sec:lbt})
was used previously by Even et al.~\cite{even+is:timetable} for scheduling algorithms.

At a high level, the $k$-AV problem is similar to, but not the same as, the graph bandwidth
problem (GBW), which is defined as follows.  Given a graph $G$ and a
positive integer $k$, decide whether it is possible to arrange the
vertices of $G$ on a line such that any two adjacent vertices in $G$
are separated by at most $k-1$ vertices on the line.  For arbitrary
$k$, where $k$ can change with the problem size, GBW is
NP-complete~\cite{papa:graph-bw} and remains so even for special kinds
of graphs~\cite{ggjk:graph-bw}.  For fixed $k$, Garey et
al.~\cite{ggjk:graph-bw} show that it is in P for $k=2$ and
Saxe~\cite{saxe:graph-bw} shows that it is still in P for $k \geq 3$.
Saxe's algorithm runs in time $O(n^{k+1})$.  Unfortunately, the
special insight exploited by Saxe~\cite{saxe:graph-bw} does not hold
for the $k$-AV problem.  When restricted to interval graphs, however,
GBW can be solved efficiently in both $n$ and $k$.  Kleitman and
Vohra~\cite{kv:gbw-interval} present an algorithm that runs in time
$O(n \log n)$ time, where $n$ is the number of vertices in the graph.
GBW is one variation of graph layout problems.  Cohen et
al.~\cite{cohen+etal:ola} show that, if the metric is to minimize the
sum (not the maximum) of the differences between the positions of two
adjacent vertices, then the problem, called optimal linear arrangement
(OLA), is NP-hard even on interval graphs.  Furthermore, GBW is
fixed-constant tractable, but otherwise NP-complete.

% LocalWords:  GBW vertices et al Kleitman Vohra Diaz LBT

% !TEX root = main.tex
% !TEX spellcheck = en_US

\section{Concluding remarks}
\label{sec:conclude}

In this paper, we made considerable progress towards resolving the
$k$-AV problem.  The primary open question that remains is to solve the $k$-AV
problem for a fixed constant $k \geq 3$, or else show that it is NP-complete.
%We have attempted to extend LBT or FZF to $k=3$ but have not been successful.
%So far the clues we have do not strongly suggest one way or the other.
Secondly, it would be interesting to test whether existing storage systems provide 2-atomicity in practice,
and understand when and why they might fail to do so.

\medskip

\noindent \textit{Acknowledgment}\ \\
We are grateful to Bob Tarjan and Steve Uurtamo for stimulating
discussions on topics related to this paper.

\bibliographystyle{plain}
\bibliography{refs}

% \appendix

% \input{proof-lbt}
% \input{proof-fzf}
% \input{proof-gpo}
% \input{proof-k-wav}

\end{document}